\documentclass[final,5p,times,twocolumn]{elsarticle}

\usepackage{amssymb}
\usepackage{epsfig}
\usepackage{graphicx}
\usepackage{amsmath}
\usepackage{amssymb}
\usepackage{booktabs}
\usepackage{multirow}
\usepackage{hhline}
\usepackage{mismath}
\usepackage{threeparttable}
\usepackage[table]{xcolor}
\usepackage[pagebackref,breaklinks,colorlinks]{hyperref}
\usepackage{subcaption}
\usepackage{wrapfig}
\usepackage{graphicx}
\usepackage{parskip} 

\usepackage[capitalize]{cleveref}
\crefname{section}{Sec.}{Secs.}
\Crefname{section}{Section}{Sections}
\Crefname{table}{Table}{Tables}
\crefname{table}{Tab.}{Tabs.}

\newcommand{\etal}{\textit{et al.}}
\journal{Neurocomputing}

\begin{document}
\definecolor{battleshipgrey}{rgb}{0.52, 0.52, 0.51}

\begin{frontmatter}


 \author[label1]{Salim Rukhsar\corref{cor1}}
 \ead{rukhsar.1@iitj.ac.in}
 \ead[url]{https://github.com/Salim-Lysiun}
 \cortext[cor1]{Corresponding Author}

 \author[label1]{Anil K.Tiwari}
 
 \affiliation[label1]{organization={Department of Electrical Engineering, IIT Jodhpur},
             postcode={342037},
             country={India}}

\title{ARNN: Attentive Recurrent Neural Network for Multi-channel EEG Signals to Identify Epileptic Seizures}


\begin{abstract}
\vspace{-10mm}

Electroencephalography (EEG) is a widely used tool for diagnosing brain disorders due to its high temporal resolution, non-invasive nature, and affordability. Manual analysis of EEG is labor-intensive and requires expertise, making automatic EEG interpretation crucial for reducing workload and accurately assessing seizures. In epilepsy diagnosis, prolonged EEG monitoring generates extensive data, often spanning hours, days, or even weeks. While machine learning techniques for automatic EEG interpretation have advanced significantly in recent decades, there remains a gap in its ability to efficiently analyze large datasets with a balance of accuracy and computational efficiency. To address the challenges mentioned above, an Attention Recurrent Neural Network (ARNN) is proposed that can process a large amount of data efficiently and accurately. This ARNN cell recurrently applies attention layers along a sequence and has linear complexity with the sequence length and leverages parallel computation by processing multi-channel EEG signals rather than single-channel signals. In this architecture, the attention layer is a computational unit that efficiently applies self-attention and cross-attention mechanisms to compute a recurrent function over a wide number of state vectors and input signals. This framework is inspired in part by the attention layer and long short-term memory (LSTM) cells, but it scales this typical cell up by several orders to parallelize for multi-channel EEG signals. It inherits the advantages of attention layers and LSTM gate while avoiding their respective drawbacks. The model’s effectiveness is evaluated through extensive experiments with heterogeneous datasets, including the CHB-MIT and UPenn and Mayo’s Clinic datasets. The empirical findings suggest that the proposed model outperforms baseline methods such as LSTM, Vision Transformer (ViT), Compact Convolution Transformer (CCT), and R-Transformer (RT), showcasing superior performance with faster processing capabilities across EEG datasets. The code has been made publicly accessible at \url{https://github.com/Salim-Lysiun/ARNN}.

\end{abstract}





\end{frontmatter}


\section{Introduction}
\label{sec:intro}
Epileptic seizures are a common neurological condition that has affected around $50$ million individuals globally \cite{Authors36,Authors9}. These seizures are marked by abrupt and abnormal electrical activity in the brain, resulting in recurring and unprovoked episodes that impede normal functioning of human beings. The occurrence and frequency of epilepsy are influenced by factors such as age, genetics, and environmental triggers. A notable increase in epilepsy cases is observed during childhood and later in adulthood \cite{Authors38}. Seizures result from a disruption in the balance between excitatory and inhibitory brain signals, caused by mechanisms like ion channel dysfunctions, neurotransmitter imbalances, and structural abnormalities, particularly in regions like the hippocampus \cite{Authors36, Authors39}. The detection of these seizures is of utmost importance, especially through the use of scalp electroencephalography (EEG), which continues to be a fundamental non-invasive method for monitoring the electrical activity of the brain \cite{Authors37}. Advancements in EEG signal processing have significantly enhanced seizure detection and its analysis capability, providing deeper insights of their dynamical behavior and thus aiding in accurate diagnosis. Effective seizure detection using scalp EEG is essential for timely intervention and management. Prolonged monitoring in epilepsy cases generates large amounts of EEG data due to extended brain activity recording \cite{Authors41}. Therefore, automatic interpretation of EEG signals is crucial to alleviate the workload by processing large amount of data to detect seizure accurately and efficiently. Advancement in machine learning techniques has the potential for efficient and personalized treatments.

\begin{figure*}
    \centering
    \includegraphics[width=0.65\textwidth, height=0.43\textwidth]{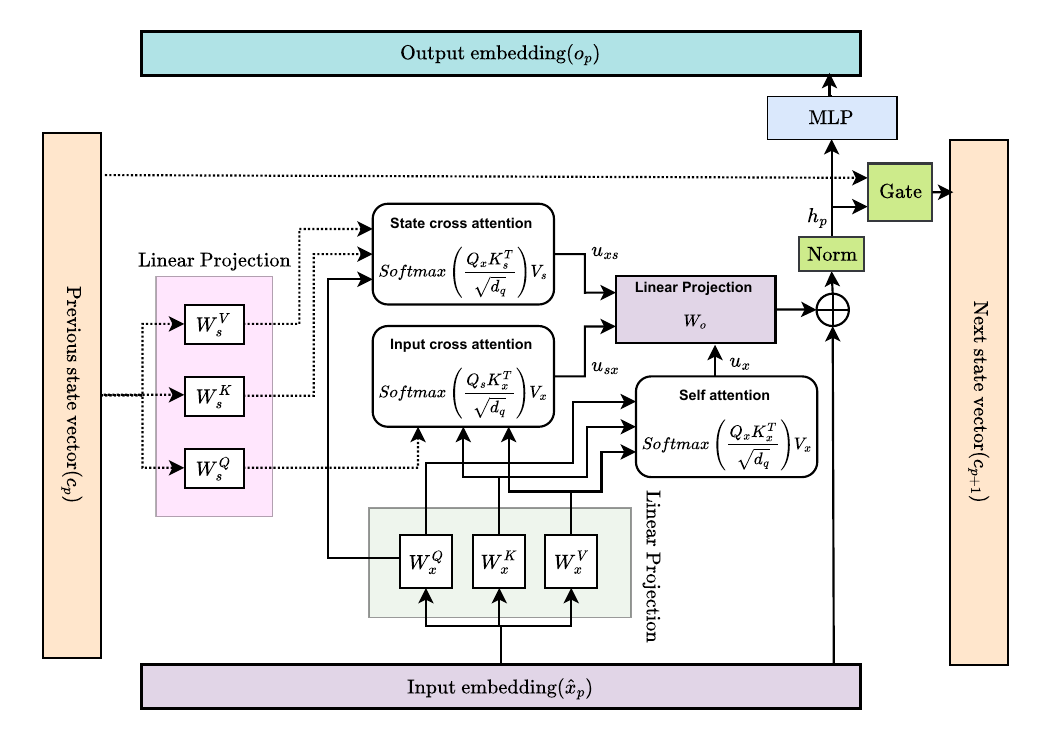}
    \caption{Comprehensive illustration of the Attentive recurrent neural network (ARNN) architecture. The architecture begins by segmenting multi-channel input EEG signals into number of local windows, enabling efficient capture of short-term dependencies. Each window undergoes linear projection to generate query (Q), key (K), and value (V) vectors, serving as inputs for the self-attention layer, which captures intra-window dependencies. The input cross-attention mechanism establishes a connection between the current input vectors and recurrent states, facilitating context-aware feature refinement. The state cross-attention layer further maintains interdependencies among state vectors across multiple windows, ensuring temporal continuity. These attention outputs are concatenated and transformed via a linear projection layer to produce a hidden state. This hidden state is processed by the LSTM-style recurrent gate, which dynamically controls the integration of current and past states, enhancing the model's ability to capture both local and global dependencies.}
    \label{fig1}
\end{figure*}

Nowadays, Transformers are preferred over recurrent neural networks (RNNs) ~\cite{Authors1} for processing non-temporal data, such as in natural language processing ~\cite{Authors3} and image recognition ~\cite{Authors4}. They have been highly successful for a number of reasons. To begin with, training Transformers is more efficient on today’s accelerator technology since all sequences are processed in parallel. RNNs excel in handling time series datasets due to their ability to capture temporal dependencies and sequential patterns, but it processes sequences one by one and thus longer training time is required. Second, while passing from one sequence to the next, an RNN has to condense the preceding sequences into a single hidden state vector. However, the amount of previous sequences that the long short-term memory (LSTM) can encode, is constrained by the size of the state vector. Transformer, on the other hand, is not restricted in its ability to process past sequences. Third, attention is efficient even across longer distances. The information may be lost by the forget gate in an LSTM when moving forward, leading to vanishing gradients during backpropagation. In practice, this implies that LSTM faces problems in sending enriched information over a large number of sequences, which is significantly less than the handling ability of the attention mechanism of the Transformer ~\cite{Authors2}. 

Despite the stated benefits, Transformers do have drawbacks. Long-form document processing, such as technical articles or ultra-high-pixel images, is hampered by the quadratic computational complexity of self-attention with respect to the sequence length ~\cite{Authors9, Authors3, Authors24}. Original Transformer architecture inherently ~\cite{Authors3} does not understand or handle relationships and interactions between different channels or variables in multichannel time series data. Furthermore, the self-attention does not find effective local dependencies and hence loses information in time series data that consists of local high frequency components ~\cite{Authors24}.

This paper introduces a novel framework that leverages both attention and recurrence, as shown in \cref{fig1}. The aforementioned restrictions are all addressed in our implementation of the Attentive recurrent neural networks (ARNN), which is unique in several key respects. This proposed model is designed to take the benefit of attention and recurrence while avoiding their respective drawbacks for processing large dataset efficiently and accurately. In this novel architecture, we combined self and cross-attention for capturing local features in a local window, and a recurrence gate to condense the information to form a global feature.
Instead of sequentially processing individual sequences, the recurrent cell functions on the local window of a segment, as illustrated in \cref{fig2}. Within each local window, all sequences are processed simultaneously. Similarly, the recurrent cell operates on a block of state vectors, rather than just a single vector. As a result, the recurrent state size is significantly larger than that of a typical LSTM, greatly enhancing the model's capacity to retain past information. The use of local window-based sequence processing allows the model to propagate information and gradients over longer distances, due to the reduced number of recurrent steps. This approach decreases the number of forget gate applications, thereby reducing training time. The model is designed to process local windows sequentially, benefiting from both attention and recurrence mechanisms. However, a trade-off has been made between these two processes to determine the optimal size of the local windows for computational efficiency. This balance ensures that the model can effectively utilize both mechanisms for efficient and accurate processing. The multi-channel EEG signals exhibit dominantly high-frequency bursts with long-term relations in seizure classification ~\cite{Authors9, Authors25}. By capturing these high-frequency components within local windows, the proposed ARNN cell can form global features that provide a rich representation of the input data. This enhanced representation leads to improved seizure detection performance. The main contributions of the proposed work are summarized as follows:

\begin{figure*}[t]
    \centering
    \includegraphics[width=1\textwidth, height=0.35\textwidth]{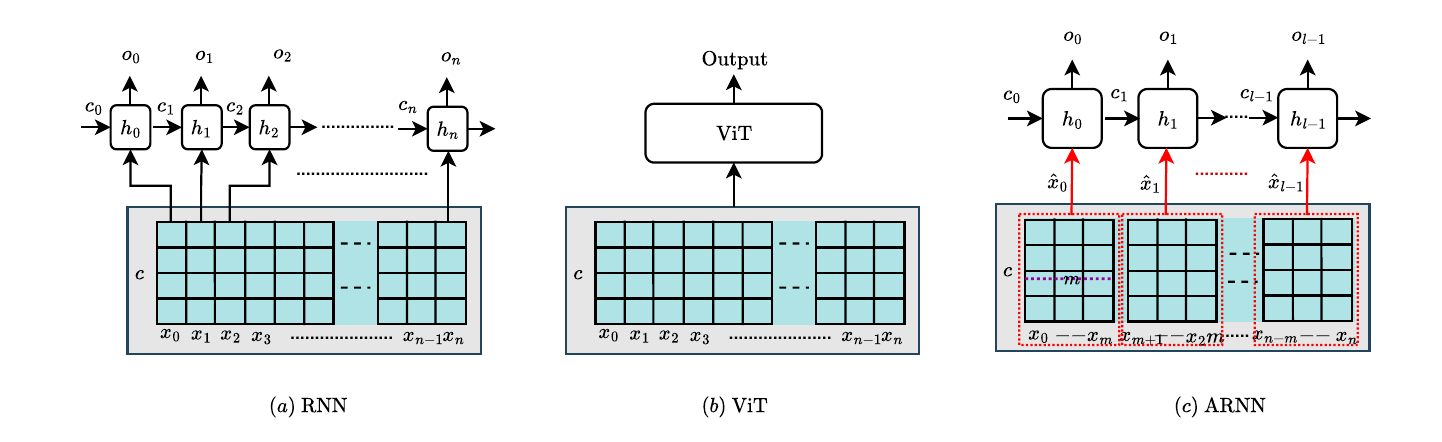}
    \caption{An illustration of sequence processing by RNN, ViT and ARNN cell. In contrast to sequential processing by RNN and parallel processing by ViT, ARNN cell operates sequentially on multi-dimensional local window of the sequence to produce global dependencies formed by local dependencies.}
    \label{fig2}
\end{figure*}

\begin{itemize}
  \item \textbf{Integration of Attention and Recurrence:} The ARNN framework uniquely combines self-attention and cross-attention mechanisms with a recurrence gate to capture fine-grained dependencies within local windows. This integration allows the model to form global features based on these dependencies, enhancing its ability to retain and process past information effectively over longer sequences.
  
  \item \textbf{Ability to process large dataset efficiently and effectively:} The proposed architecture is designed to handle large datasets by optimizing computational efficiency through the use of local window-based sequence processing. This approach reduces the number of recurrent steps and leverages attention mechanisms, enabling the model to process large amount of data more quickly while maintaining high accuracy.
  
  \item \textbf{Robust Performance in Seizure Detection:} The ARNN framework demonstrates robust performance in seizure detection by effectively capturing both high-frequency bursts and long-term dependencies within EEG signals. By leveraging the attention-recurrence mechanism, the model achieves superior accuracy in identifying seizure patterns across varying signal lengths and noise levels, making it well-suited for real-world applications in medical diagnostics.
\end{itemize}

The rest of the paper is organized as follows: Section 2 is dedicated to related work. The proposed architecture of the ARNN model is elaborated in detail in Section 3. The ablation study is presented in Section 4. The experiments and results are provided in Sections 5 and 6, respectively. Finally, Section 7 outlines the concluding remarks. 

\section{Related  Work}

Seizure detection using EEG signals has witnessed remarkable advancements with the application of deep learning techniques ~\cite{Authors5, Authors14, Authors17, Authors33}. In the recent years, numerous studies have shown the effectiveness of the sophisticated neural network architectures such as LSTM networks ~\cite{Authors24, Authors26, Authors11}, and Transformers ~\cite{Authors9, Authors10} for enhancing the accuracy and reliability of seizure detection and prediction. This section presents an overview of the recent literature that exemplifies the contributions made in this direction.

LSTM networks ~\cite{Authors22, Authors8, Authors7} have shown substantial promise in capturing temporal dependencies within EEG signals, making them well-suited for seizure detection and prediction tasks. Hezam \etal ~\cite{Authors5} proposed a novel hybrid LSTM model that leverages both short-term and long-term EEG patterns to achieve enhanced seizure detection accuracy. Similarly, the authors ~\cite{Authors6} introduced a stacked LSTM architecture to capture complex relationships in EEG signals and reported improved seizure prediction performance. Another approach involves a Generative Adversarial Network (GAN) model to generate preictal data, followed by LSTM for classification to predict outcomes ~\cite{Authors7}. Furthermore, research has extended beyond traditional LSTMs to their variants like Gated Recurrent Units (GRUs)  ~\cite{Authors8}. A comparative study in ~\cite{Authors26}, has shown its supriority over of LSTMs in terms of computational efficiency, while maintaining competitive performance in seizure detection tasks.

Recent advancements in the application of Transformers to sequential data have been extended to EEG analysis for seizure detection and predictions. Recently, a lightweight Transformer architecture has been proposed to enhance accuracy in detecting seizure patterns by capturing local and global dependencies within EEG signals ~\cite{Authors9}. Similarly, ~\cite{Authors10} introduced a Multi-channel Vision Transformer (MViT) for automated spatiotemporal spectral features learning in multi-channel EEG data to achieve state-of-the-art performance in seizure prediction. Researchers have also explored hybrid architectures that combine the strengths of LSTM and self-attention for classification and prediction from time-series data ~\cite{Authors11}. A recent study ~\cite{Authors12} introduced a hybrid model that employs a 1-D Convolutional Neural Network (CNN) with a Bidirectional Long Short-Term Memory (BiLSTM) network to detect seizures. Convolutional tokenizer is combined with ViT to effectively capture the temporal and spatial information in time series dataset ~\cite{Authors9, Authors32}. In another work ~\cite{Authors27}, R-Transformer is proposed where they used LocalRNN for local dependencies in series with a self-attention mechanism to capture global dependencies. The R-Transformer used in their work enjoys both the abilities; however, it demands more computational resources as compared to the proposed ARNN model. In implementing the proposed architecture, the training time is significantly reduced due to the recurrent use of the attention mechanism compared to other baseline methods.

\section{Attentive Recurrent Neural Networks (ARNN)}
The architectural framework of the proposed ARNN is comprehensively described in \cref{fig1}. This architecture is constructed for multichannel EEG signals that can harness the advantage of attention mechanism and LSTM-style recurrent gate, organized hierarchically. \cref{fig1} provides a detailed overview of the ARNN  architecture, specifically designed to capture both local and global dependencies in multi-channel EEG signals for seizure detection. The model begins by processing multi-channel EEG input signals, which are segmented into sequences of local windows to focus on short-term dependencies. A linear projection layer then transforms these segments into query $(Q)$, key $(K)$, and value $(V)$ vectors, forming the basis for the attention mechanisms. The self-attention mechanism computes the relationships between all positions within a local window, enhancing the representation of localized features. In parallel, the input cross-attention mechanism captures interrelationships between the input sequences and the recurrent states, allowing the model to refine its feature extraction by comparing current inputs with past information. The state cross-attention layer focuses on interactions among the state vectors themselves, preserving temporal dependencies across longer sequences. These outputs are concatenated and passed through a final linear projection, producing the hidden state, which is then processed by the recurrent gate that manages the flow of information by deciding which data to retain or discard. This gate facilitates the integration of local dependencies into global feature representations, enhancing the model's capacity to capture complex patterns in EEG data. The combination of attention layers and recurrent gating enables ARNN to efficiently handle extensive EEG datasets, leveraging both the local characteristics of short bursts and the long-term dependencies of sequential patterns.

\subsection{ARNN cell}
The process of handling data segments in a single ARNN cell, is explained in this section. Let a single multi-channel EEG segment $\mathcal{X} = x_{t}|_{t=0}^{n}= \{x_0, x_1, x_2, x_3, \dotsb x_{n}\} \in \mathbb{R}^{c\times n}$ where $x_{t}\in \mathbb{R}^{c}$, $c$ denotes the number of channels and $n$ is the sequence length. The goal is to have a function that can transform the input EEG segments into a labels space $\mathcal{Y}: (f:\mathcal{X}^{c\times n} \rightarrow \mathcal{Y})$. This can be formulated as below,
\begin{equation*}
    y = f(x_0, x_1, \dotsb x_n)
\end{equation*}
where is $y\in \mathcal{Y}$ is the seizure and non-seizure labels. EEG data inherently exhibits strong local structures. Therefore, it is both desired and essential to include such elements in its representation that capture these local characteristics. The proposed architecture is designed to meet this requirement by capturing local short-term dependencies, and also managing them to form global dependencies. 

In contrast to other efforts that typically apply RNNs to the entire sequence ~\cite{Authors33, Authors22, Authors7}, we reconstruct the given long sequence of $n$ samples into numerous shorter ones, each of which is expected to contain local information and is processed by an ARNN cell. Specifically, $l$ local windows of $m$ samples were created as follows: $\hat{x}_p= \{x_{mp+1}, x_{mp+2}, \dotsb x_{(p+1)m}\}$ of $m$ consecutive points forms $p= [0, 1, \dotsb l-1]$ local windows. The ARNN cell will learn a latent representation from the points in each local window $\hat{x}_p$, and recurrently manage to form global long-term dependencies of the segment $x_t$. Thus, the learned latent representations contain the local structure information of the long sequence. Therefore, the ARNN cell forms global dependencies by focusing on the local dependencies of local windows. \cref{fig2} shows the key difference in processing a segment by an RNN, ViT, and the proposed ARNN cell.  

\begin{figure}
    \centering
    \includegraphics[width=\linewidth]{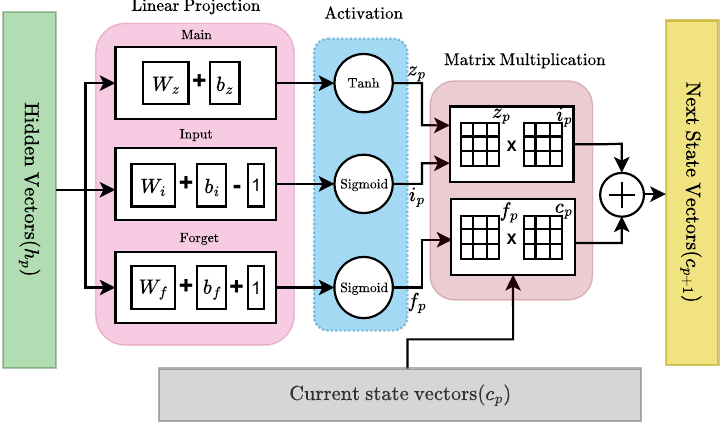}
    \caption{Comprehensively description of LSTM-style recurrent gate architecture designed in ARNN cell.}
    \label{fig3}
\end{figure}

\begin{figure*}[t]
\begin{subfigure}{0.5\textwidth}
\includegraphics[width=1\linewidth, height=4.5 cm]{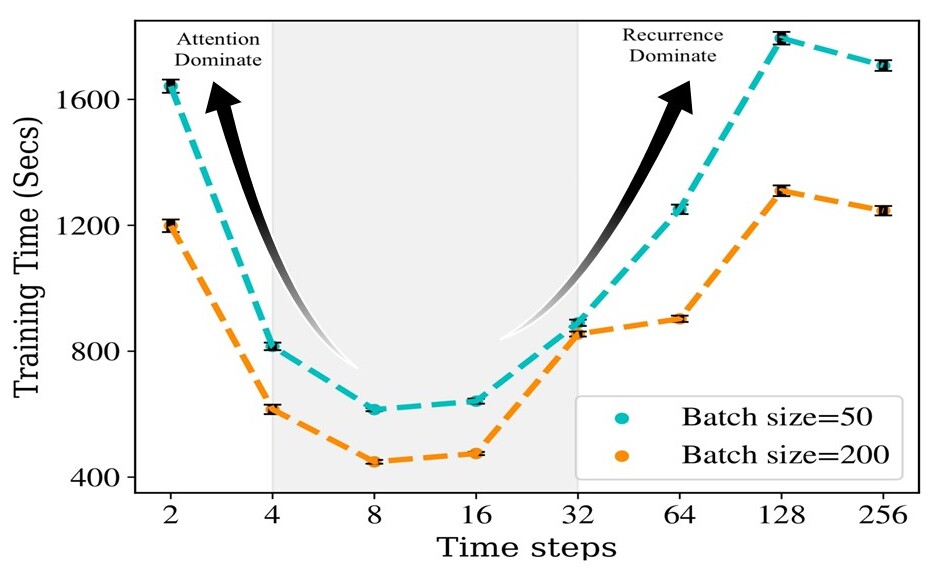} 
\caption{Time steps of the ARNN cell versus training time at batch size 50 and 200.}
\label{fig6a}
\end{subfigure}
\hfill
\begin{subfigure}{0.5\textwidth}
\includegraphics[width=1\linewidth, height=4.8cm]{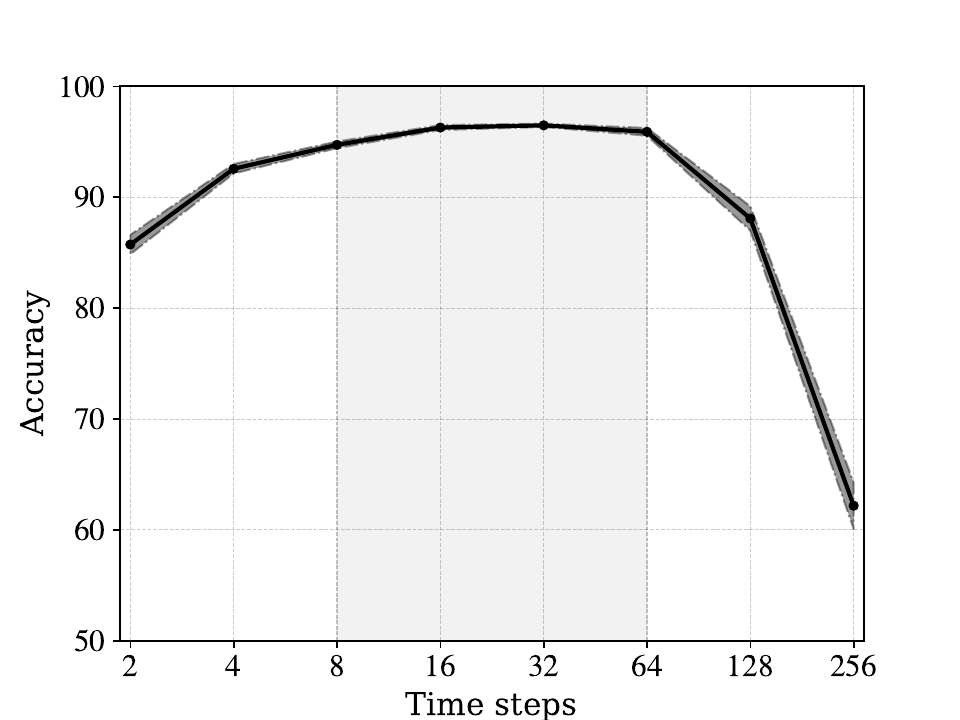}
\caption{Time steps versus accuracy of the ARNN cell.}
\label{fig6b}
\end{subfigure}
\caption{An illustration depicting the effect of the time step on the (a) training time and (b) accuracy of the ARNN cell at the segment length of 1024 of multi-channel EEG signals. The grey shadow indicates the optimal time step for the cell.}
\label{fig6}
\end{figure*}

\textbf{Attention layer:}
Recent studies indicate that the multi-head attention mechanism is highly effective in capturing global dependencies, by enabling direct connections between all pairs of positions ~\cite{Authors12, Authors32, Authors27}. In this attention mechanism, each position attends to preceding positions, generating a set of attention scores to refine its representation. Mathematically, the current local window $\hat{x}_p$ can be processed as follows:

\begin{equation}
u_x = \textit{softmax}\left(\frac{Q_xK_x^T}{\sqrt{d_k}}\right)V_x 
    \label{eqn1}
\end{equation}

Where $Q_x$, $K_x$ and $V_x$ are the query, key and value vectors of $\hat{x}_p$, derived from the linear projection of $W_x^Q$, $W_x^K$ and $W_x^V$ weight matrices, and $d_k$ is the dimension of key vectors. The scaling factor $\sqrt{d_k}$ is used to  prevent the dot product of the query and key vectors from growing too large, which can negatively affect the performance of the softmax function, causing it to become extremely small and lead to vanishing gradients.
The cross-attention mechanism enables the model to capture the interrelation and temporal dependencies between the input vector $\hat{x}_p$ and the state $c_p$, thereby focusing on crucial local position vectors. This mechanism is effective in identifying seizures accurately by capturing interrelations across different sequences. Specifically, in cross-attention mechanisms, each element in one sequence attends to all elements in another sequence, generating attention scores crucial for refining its representation. The interrelation between the input vector $\hat{x}_p$ and the state vectors $c_p$ of the ARNN cell is formulated in \cref{eqn2}:

 \begin{equation} 
	\begin{split}
	 u_{xs} & = \textit{softmax}\left(\frac{Q_xK_s^T}{\sqrt{d_k}}\right)V_s \\
	u_{sx} & = \textit{softmax}\left(\frac{Q_sK_x^T}{\sqrt{d_k}}\right)V_x ,
	\end{split}\label{eqn2}
\end{equation}
Where $Q_s$, $K_s$ and $V_s$ are the state query, key and value vectors of $c_p$ derived from a linear projection of $W_s^Q$, $W_s^K$ and $W_s^V$ weight matrices, as shown in \cref{fig1}. The hidden state of the cell is computed by the linear projection of concatenated vectors from $u_x$, $u_{xs}$ and $u_{sx}$ as in \cref{eqn3}
\begin{equation}\label{eqn3}
	 h_p = [u_x\hspace{2pt} u_{xs} \hspace{2pt} u_{sx}]^T W_o
\end{equation}

This hidden state $h_p$ caries the information learned from the current input $\hat{x}_p$ and the previous state vectors $c_p$. The normalized hidden state is used by the gate to condense the information and carry to the next state.

\textbf{Recurrent Gate:}
We used the LSTM-style recurrent cell for gating purposes. This gate uses the standard combination of input and forget gates as shown in \cref{fig3}. Although there is a separate gate for each state vector, they are allowed for their simultaneous updating. The following equations are used for the recurrent gate:
\begin{equation}
\begin{split}
    z_p = tanh(W_zh_p + b_z)\\
    i_p = \sigma(W_ih_p + b_i -1)\\
    f_p = \sigma(W_fh_p + b_f + 1)\\
    c_{p+1} = c_p\odot f_p + z_p \odot i_p\\
\end{split}\label{eqn4}
\end{equation}

The trainable weight matrices $W_z$, $W_i$, and $W_f$, and the trainable bias vectors $b_z$, $b_i$, and $b_f$ are essential components of the recurrent gate. The LSTM-style gate demonstrates greater expressive power due to the dependency of $f_p$ and $i_p$ values on the current hidden state vectors $h_p$ as shown in Fig. \ref{fig3} and expressed in Eq. \ref{eqn4}. In the model, $h_p$ depends on $c_p$, thereby inducing an indirect dependency of the LSTM gate on $c_p$. Consequently, recurrent gate values vary across each state vector and block indices $p$. We incorporate a slight adjustment of $-1$ and $+1$ to the input and forget gates. This adjustment serves to initially influence the gate towards "remembering" while maintaining the scale of updates. This initialization technique enables the recurrent cell to consistently leverage the recurrent state.

\subsection{Complexity Analysis}
In the following section, we compared the time complexity of proposed ARNN cell with self-attention mechanism and analyze its computational speed.

\textbf{Attention mechanism of Transformer:} Consider the \cref{eqn1}. For a segment \(\mathcal{X}\) of dimension  \( (c, n) \), computing the output of self-attention requires the following steps.

\begin{itemize}
    \item The segment \(\mathcal{X}\) is linearly transformed to generate the query \(Q\), key \(K\), and value \(V\) matrices, each with a dimension of \((c, n)\). This transformation is achieved by multiplying \(\mathcal{X}\) with three learned matrices, each of dimension \((n, n)\), resulting in a computational complexity of \(\mathcal{O}(c n^2)\).
    \item To compute the layer output as described in \cref{eqn1}, the formula \(\textit{softmax}\left(\frac{QK^T}{\sqrt{d_k}}\right)V\) is used. The computation of \(QK^T\) has a complexity of \(\mathcal{O}(c^2 n)\), and multiplying the resulting matrix by \(V\) also has a complexity of \(\mathcal{O}(c^2 n)\).
\end{itemize}
Therefore, the total complexity of the self-attention mechanism is \(\mathcal{O}(c n^2 + c^2 n)\).

\textbf{ARNN self-attention.} In the ARNN cell, the segment \(\mathcal{X}\) is divided into \(l\) local windows ,\(\hat{x}_p\), of length \(m\), where \(m = \frac{n}{l}\). The complexity of the self-attention mechanism in the ARNN cell can be described as follows:
\begin{itemize}
    \item The linear transformation of a local window \(\hat{x}_p\) to generate the \(Q\), \(K\), and \(V\) matrices, each with a dimension of \((c, m)\), requires a computational complexity of \(\mathcal{O}\left(c \left(\frac{n}{l}\right)^2\right)\).
    \item The computation of \(QK^T\) for a local window has a complexity of \(\mathcal{O}\left(c^2 \frac{n}{l}\right)\), and multiplying this result by \(V\) has a complexity of \(\mathcal{O}\left(c^2 \frac{n}{l}\right)\).
\end{itemize}

For the ARNN cell to process a segment \(\mathcal{X}\) of dimension \( (c, n) \), the total computational complexity is \( \bigO \left(\frac{c \cdot n^2}{l} + \frac{c^2 \cdot n}{l}\right) \). The number of local windows, \( l \), plays a crucial role in this process. A larger \( l \) causes the ARNN cell to behave more like an RNN, while a smaller \( l \) makes it rely primarily on the attention mechanism, as illustrated in Fig. 4a. Therefore, a trade-off between recurrence and attention must be achieved to leverage the benefits of both. The ARNN cell reduces space and time complexity by reducing the complexity of self-attention from \(\mathcal{O}(c n^2 + c^2 n)\) in the standard attention mechanism to  \( \bigO \left(\frac{c \cdot n^2}{l} + \frac{c^2 \cdot n}{l}\right) \). By processing smaller windows, the ARNN cell reduces computational cost and memory usage, enhancing the model's efficiency compared to traditional attention mechanisms while effectively balancing the benefits of recurrence and attention. Fig. \ref{fig7b} and \ref{fig7c} illustrate the comparative training and inference times of the ARNN model compared to baseline models across different sequence lengths.

\section{Experiments}
\textbf{Set-up:} Experiments were run on a computer with an Apple M1 processor with 16 GB RAM, running on macOS Ventura version 13.4.1(c). The chosen batch size for training is 50, offering a balance between computational efficiency and gradient accuracy. Initially, the learning rate is set to $0.001$, ensuring stable optimization of model parameters. But this rate is decreased by a factor of $0.1$ at every $10$ epochs to fine-tune the model convergence. A Dropout layer with a dropout rate of \(0.3\) is used to regularize the model. The model is trained for 30 epochs using the Adam optimizer, chosen for its efficiency in handling sparse gradients and noisy data. Binary cross-entropy is utilized as the loss function, providing a measure of the discrepancy between predicted and actual labels. The effect of batch size is shown in shown in Fig. \ref{fig6a}. The data sets were divided into $75\%$ for training and $25\%$ for testing. However, the validation is not performed due to the small size of the datasets. The test result averages were calculated after ten iterations of each model.

\begin{table*}[h]
  \centering
  \rowcolors{4}{gray!5}{gray!15}
  \begin{threeparttable}
  \begin{tabular}{c |c c c c c c c c c c}
    \hline
    \rowcolor{battleshipgrey!80}
     & \multicolumn{2}{c|}{LSTM} &\multicolumn{2}{|c|}{ViT} &\multicolumn{2}{|c|}{CCT} &\multicolumn{2}{|c|}{RT} &\multicolumn{2}{|c}{ARNN} \\
     \cline{2-11}
     \rowcolor{battleshipgrey!80}
      \multirow{-2}{*}{Participants}  & \cellcolor{battleshipgrey!40}Acc & \cellcolor{battleshipgrey!40}f1-score & \cellcolor{battleshipgrey!40}Acc & \cellcolor{battleshipgrey!40}f1-score & \cellcolor{battleshipgrey!40}Acc & \cellcolor{battleshipgrey!40}f1-score & \cellcolor{battleshipgrey!40}Acc & \cellcolor{battleshipgrey!40}f1-score & \cellcolor{battleshipgrey!40}Acc & \cellcolor{battleshipgrey!40}f1-score \\\hline
      Dog-1 & 0.745&  0.768 &  0.718&  0.726&  0.973&  0.972& 0.971 & 0.973 & 0.979& \textbf{0.979}\\\cline{2-11}    
    Dog-2 & 0.869 &  0.917 &  0.894&  0.907& 0.981& \textbf{0.982}& 0.978& 0.978& 0.982& \textbf{0.982}\\\cline{2-11}
   Dog-3 & 0.951 &  0.953 & 0.980& 0.981& 0.990&  0.991& 0.991&  0.991&  0.995& \textbf{0.995}\\\cline{2-11}
   Dog-4 & 0.934 & 0.947 & 0.959& 0.959&  0.977& 0.978& 0.984& 0.984& 0.986&  \textbf{0.986}\\ \cline{2-11}
    \textbf{Average} & 0.874 & 0.896 & 0.888 & 0.893 & 0.980 & 0.981 & 0.981 & 0.981 & 0.985 & \textbf{0.986} \\ \hline
   Patient-1\tnote{a} & 0.590 & 0.594& 0.522 &0.520 & 0.795 &   \textbf{0.811} & 0.501 & 0.502 & 0.704 & 0.713\\\cline{2-11}
   Patient-2 & 0.983 & 0.984 & 0.987 &0.988  & 0.988 & 0.989 & 0.993 & 0.993 & 0.994 & \textbf{0.994} \\\cline{2-11}
   Patient-3 &  0.755&  0.767&  0.819&  0.821&  0.965&  \textbf{0.966} & 0.912 & 0.913 &  0.943 & 0.943\\ \cline{2-11}
   Patient-4\tnote{b} &  0.887&  0.940&  0.849&  0.863&  0.849&  0.884 & 0.905 & \textbf{0.937} &  0.869 & 0.888\\ \cline{2-11}
   Patient-5 & 0.955 & 0.971 &  0.981&  0.983&  0.992 &  \textbf{0.992} & 0.984 & 0.985 &  0.988 & 0.989\\ \cline{2-11}
   Patient-6 & 0.908 & 0.951 & 0.967 & 0.967 & 0.987 & \textbf{0.987} & 0.987 & 0.987 & 0.982 & 0.982  \\ \cline{2-11}
   Patient-7 & 0.923 & 0.944 & 0.969 & 0.969 & 0.964 & 0.963 & 0.973 & 0.974 & 0.985 & \textbf{0.985} \\ \cline{2-11}
   Patient-8 & 0.945 & 0.948 & 0.964 & 0.963 & 0.983 & 0.983 & 0.983 & 0.983 & 0.988 & \textbf{0.988} \\ \cline{2-11}
 \textbf{Average}\tnote{c} & 0.912 & 0.927 & 0.948 & 0.950 & 0.979 & \textbf{0.980} & 0.972 & 0.973 & 0.980 & \textbf{0.980}\\ \hline
  \end{tabular}
  \begin{tablenotes}\tiny
        \item[a] Excluded from average due to 174 sec available data. 
        \item[b] Excluded from average due to 210 sec available data.
        \item[c] Average performance of patients excluding 1 and 4.
    \end{tablenotes}
  \end{threeparttable}
   \caption{A comparative analysis of accuracy and f1-score among the proposed ARNN and other state-of-the-art implemented methods on UPenn $\&$ Mayo Clinic's seizure detection challenge dataset.}
   \label{Tab1}
\end{table*}

\begin{figure}[h]
    \centering
    \includegraphics[width=\linewidth, height = 4cm]{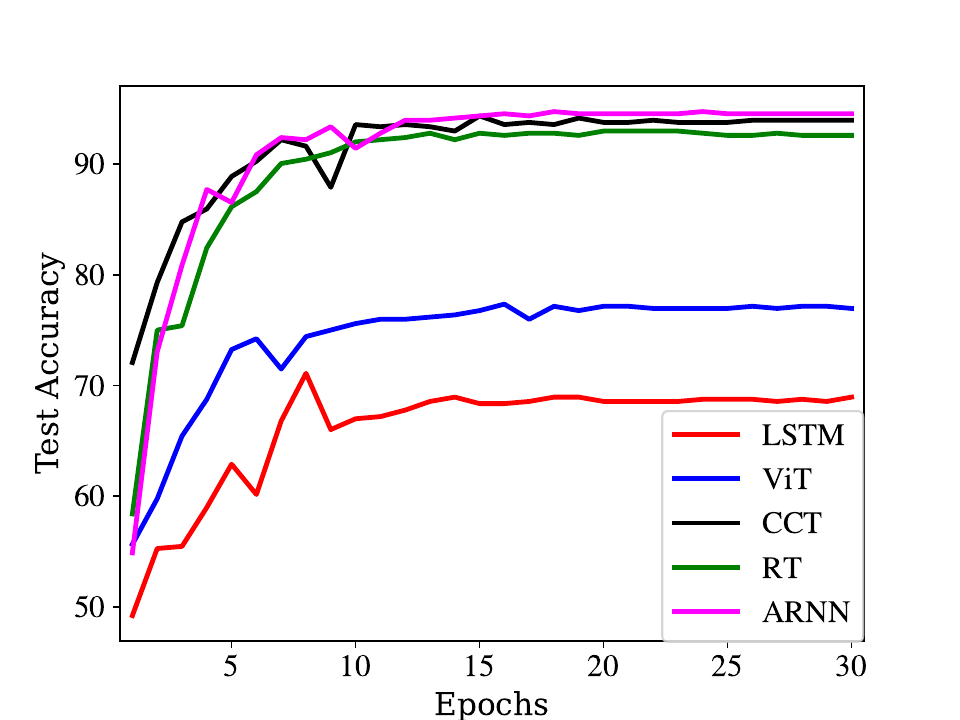}
    \caption{Testing accuracy of the proposed model against the baseline models across epochs.}
    \label{fig4}
\end{figure}

\begin{figure}[h]
  \centering
   \includegraphics[width=\linewidth, height = 4cm ]{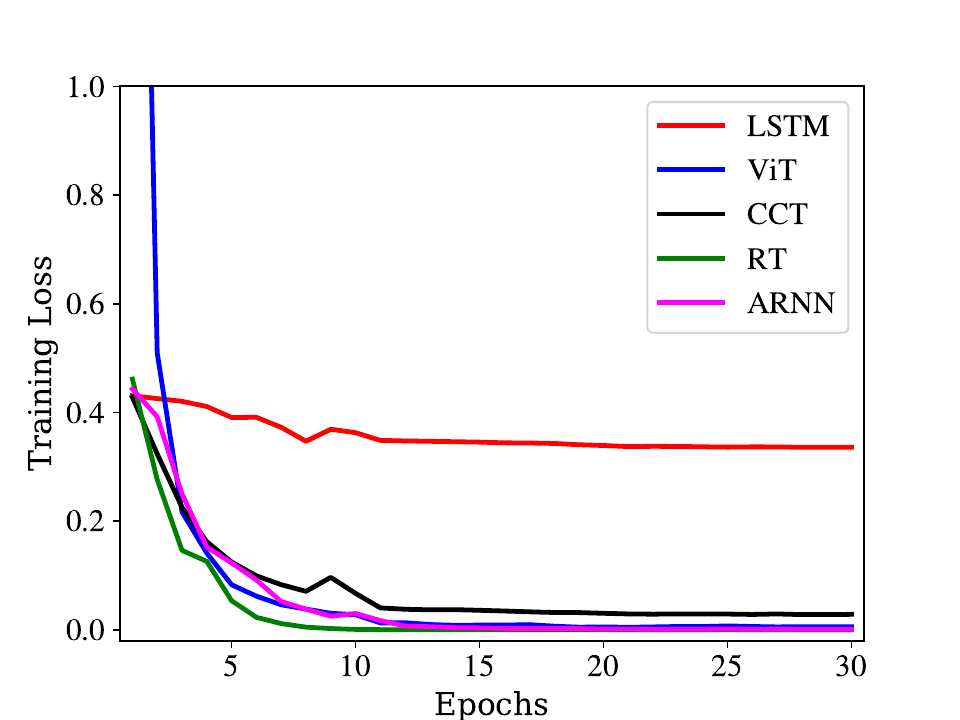}
   \caption{The evolution of training loss is compared between the proposed model and baseline models across different epochs.}
   \label{fig5}
\end{figure}

\subsection{Datasets}
\textbf{UPenn and Mayo Clinic's Seizure Detection Challenge} \footnote{\url{https://www.kaggle.com/c/seizure-detection}}: A seizure detection challenge was organized by the National Institute of Neurological Disorders and Stroke (NINDS), and the American Epilepsy Society (AES) in 2014 ~\cite{Kagle1, Authors18, Authors20}. The goal of this challenge was to facilitate the identification of specific brain regions susceptible to surgical resection in order to mitigate the occurrence of future seizures. The EEG recordings of four canines were recorded with 16 electrodes at 400 Hz, whereas human EEGs were recorded with varying electrode counts at frequency 500 or 5000 Hz. All data were divided into unbalanced 1 second EEG segments and labeled as “ictal” or “interictal”. This challenge aimed to gain essential insights into the shared characteristics and differences in epileptic neural activity across species by integrating data from diverse sources, including both animal and human subjects. The resulting perspectives hold significant value in the enhancement of clinical interventions and the formulation of more effective treatment methods. The model is evaluated on the labeled data of all the participants. All data were resampled at 400 Hz to ensure consistency for the proposed model. The empirical findings are presented in Table \ref{Tab1}. 

\begin{figure}[ht]
\begin{subfigure}{0.5\textwidth}
\includegraphics[width=0.9\linewidth, height=4cm]{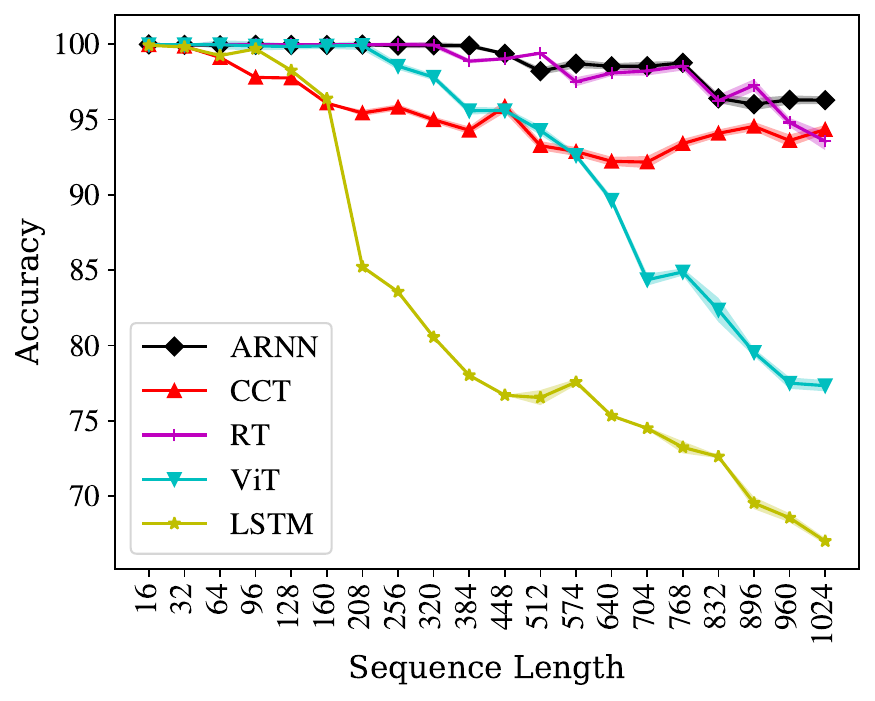} 
\caption{Accuracy against the sequence length of a segment.}
\label{fig7a}
\end{subfigure}

\begin{subfigure}{0.5\textwidth}
\includegraphics[width=1\linewidth, height=4cm]{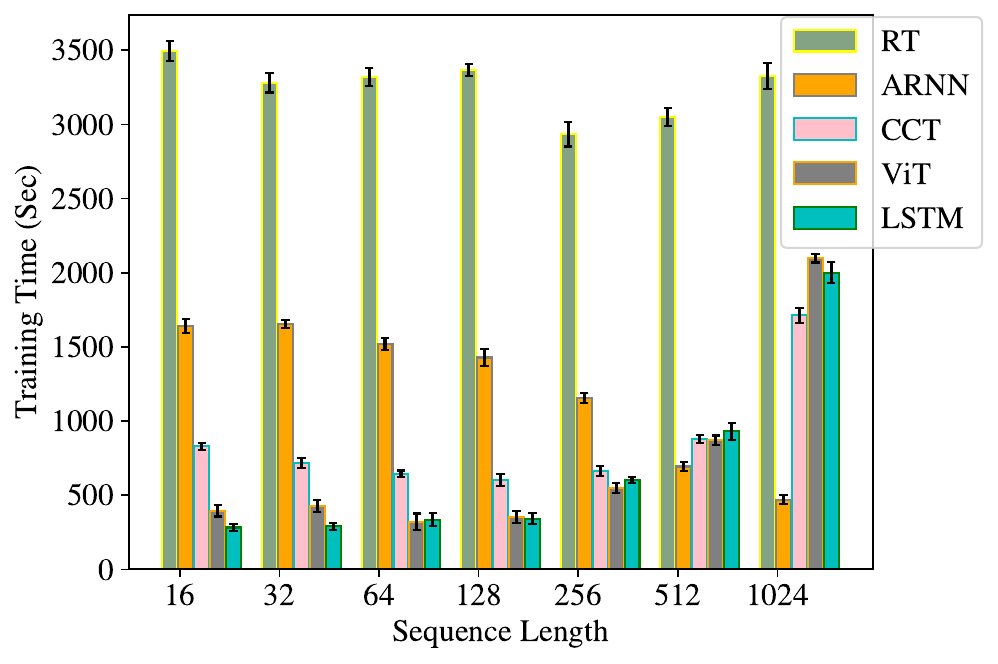}
\caption{Time required for training with respect to the sequence length of a segment.}
\label{fig7b}
\end{subfigure}

\begin{subfigure}{0.5\textwidth}
\includegraphics[width=01\linewidth, height=4cm]{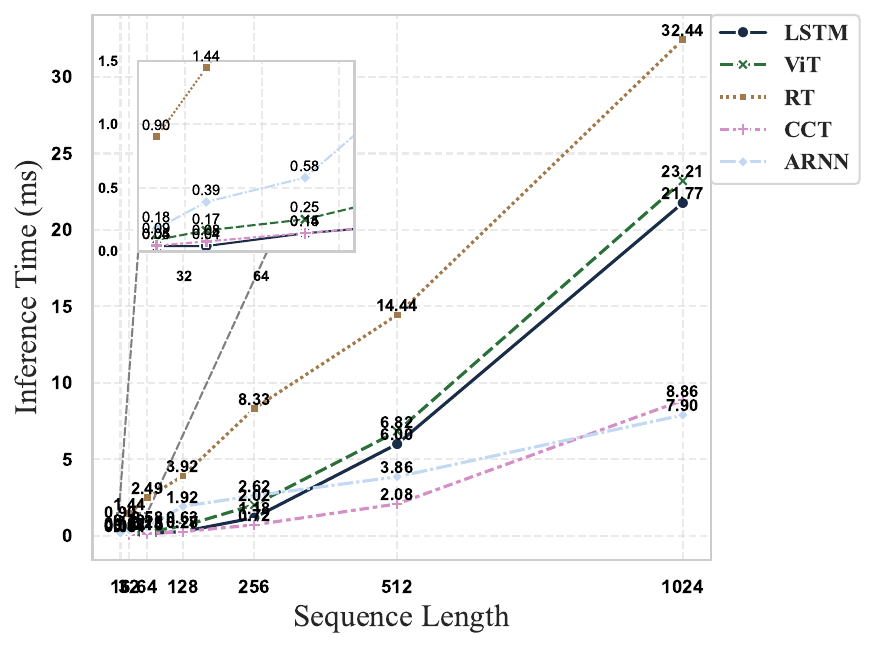}
\caption{Inference time comparison across sequence lengths.}
\label{fig7c}
\end{subfigure}

\caption{Comparative analysis of the performance of ARNN and baseline models across different sequence lengths, highlighting accuracy in (a), training time in (b), and (c) compares the inference times (ms) of various models—LSTM, ViT, RT, CCT, and ARNN. Demonstrates efficient scaling with consistent performance of ARNN, offering a favorable balance of computational efficiency and accuracy, particularly for longer sequences, compared to RT and ViT models.}
\label{fig7}
\end{figure}

\textbf{CHB-MIT}\footnote{\url{https://ieee-dataport.org/open-access/preprocessed-chb-mit-scalp-eeg-database}}: Long-term scalp EEG recordings were collected at Boston Children's Hospital and publicly released as the CHB-MIT scalp EEG dataset on \url{https://physionet.org}. This dataset was recorded from $18-23$ channels using $21$ electrodes, which were labeled according to the international $10-20$ electrode placement standard. The dataset includes recordings from $24$ pediatric patients with intractable seizure disorders, and it is widely used for seizure detection and prediction work ~\cite{Authors16}. The imbalance in this dataset is addressed by a structured approach as outlined by Deepa \etal ~\cite{Authors13}. First, the CHB-MIT EEG database from Physionet was acquired in \lq .edf\rq\ format, which includes comprehensive information about epileptic periods. The ictal and preictal states were extracted and labeled as \lq 1\rq\ and \lq 0\rq, respectively. Focus was then placed on retaining 23 essential channels out of the 96 available in the dataset and ensured the data quality by removing duplicates and incorrect electrodes. Finally, the preictal and ictal states were merged into a more balanced dataset, which helped mitigate the impact of class imbalance and improve the model's effectiveness in seizure detection. This balanced dataset comprises 68 minutes of epileptic ictal periods and an equivalent 68 minutes of preictal data. This balanced dataset is then processed by using min-max normalization method. The segmentation in this dataset differs from that of the UPenn and Mayo datasets, with segment lengths of $[0.0625, 0.125, 0.25, 0.5, 1.0, 2.0, 4.0]$ seconds. This variation is used to evaluate the ARNN model’s ability to handle different sequence lengths, as illustrated in \cref{fig7}. This pre-processed dataset was then splitted into \(75\%\) for training and \(25\%\) used to test the designed model's ability.

\section{Ablation Study}
\textbf{Time steps.}  The study was initiated with a robust baseline configuration, employing the ARNN cell at different time steps. Subsequent variations with the time steps as $2^k$ for $0 < k \leq 8$, as shown in Fig. \ref{fig6a} for $50$ and $200$ batch sizes. The experiments revealed that while a time step of $8$ offers the shortest training time, a time step of $16$, though little longer, provides $2.85\%$ higher accuracy with a p-value of $0.011$. Results from the investigation demonstrated that the ARNN model achieved an accuracy of $96.37\%$ with the less training time at $1024$ sequence length, shown in \cref{fig7a}. However, reducing the time steps from $16$ to $8$ incurred a slight decrement in accuracy to $93.52\%$, reflecting a discernible loss in feature richness, as shown in \cref{fig7b}. Further reduction to $2$ time steps accentuated this effect, yielding in drop of accuracy to $86.51\%$ and increased the training time, indicating a substantial compromise in feature retention. Conversely, extending the time steps to $64$ showcased no enhancement in accuracy to $96.35\%$, albeit with increased computational complexity. The findings emphasized the delicate balance between sequence length, computational efficiency, and detection performance in seizure detection tasks leveraging ARNN architectures. We conclude that the $16$ time step is sufficient for the model at $1024$ sequence length segment, although $8$ and $32$ time steps are preferred depending on the sequence length.\\

\begin{figure}
\includegraphics[width=1\linewidth, height=4cm]{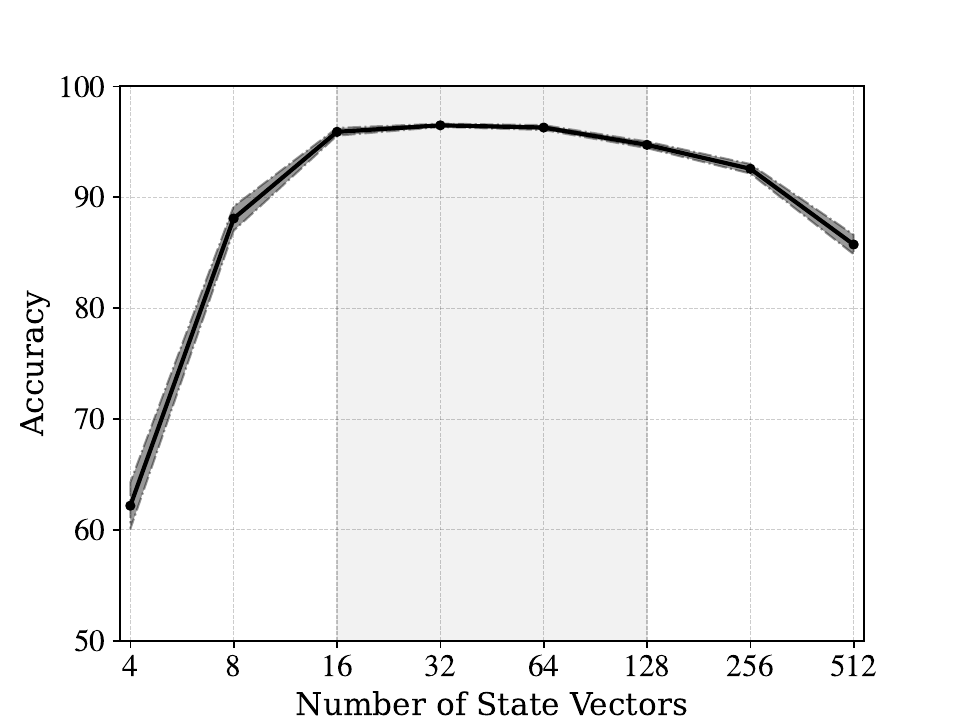}
\caption{Effect of number of State Vectors on accuracy: The plot illustrates the relationship between the number of state vectors and model accuracy.}
\label{fig8}
\end{figure}

\textbf{Number of recurrent state vectors.} The model is trained with differing numbers of state vectors, from $4$ to $512$. Initially, decreasing the number of state vectors from $512$ to $64$ increases the accuracy and decreases the computational time, but provides optimal performance in the range of $16-64$. Moreover, further decrease in the state vectors drastically decreased the accuracy and increased the training time, and worst performance with $4$ state vectors (see \cref{fig8}). This conjecture suggests that the model encounters challenges in effectively leveraging the recurrent state when the state space expands or is compressed excessively.\\
\textbf{Segment length.} Decreasing the length of the segment significantly enhanced the accuracy of the ARNN model and other baseline methods as shown in \cref{fig7a}. However at larger segment lengths, the performance of the baseline methods decreases rapidly with increase in time complexity. The ARNN performance slightly deviated but the training time decreased rapidly, as shown in \cref{fig7b}. We hypothesized that as the segment length increases, the proposed architecture summarizes the information more efficiently and effectively than the baseline methods.

\section{Results}
We tested the proposed ARNN model on two different heterogeneous datasets: Mayo Clinic’s Seizure Detection Challenge dataset ~\cite{Authors18, Authors20, Authors19} and CHB-MIT ~\cite{Authors34, Authors35}. The comparative testing accuracy on the CHB-MIT dataset, in relation to baseline methods, is shown in \cref{fig4}. Additionally, the comparative learning abilities of the models are illustrated in \cref{fig5}. The performance of the ARNN cell on these datasets is compared with both state-of-the-art methods and baseline methods.

\subsection{Performance on datasets}
\textit{Mayo Clinic’s Seizure Detection Challenge dataset:} ARNN demonstrates better performance compared to baseline models in this particular task, except slightly lag in patients 3, 5 \& 6 compared to CCT, as described in \cref{Tab1}. It is hypothesized that this phenomenon arises due to the presence of pronounced intrachannel local features within the multi-channel EEG data, and the same was captured by ARNN. Although the CCT model has a multi-head attention mechanism, which is highly efficient in capturing long-term dependencies, it lacks to capture local features of the segment that provides important feature. Furthermore, the R-Transformer has demonstrated significantly superior outcomes in the case of patients 4 \& 6 than the proposed model. Additionally, the execution time taken by each model is shown in \cref{fig7b}, further highlighting the computational efficiency of the proposed model. This demonstrates that the proposed model not only achieves better performance but also does so with greater efficiency. Moreover, it exhibits superior performance compared to the Variational Autoencoder (VAE) by a significant margin ~\cite{Authors18}. This phenomenon can be anticipated as the attention model has a tendency to disregard the sequential information present in the local structure. Fosil \etal ~\cite{Authors20} employed the discrete log energy entropy feature and support vector machine (SVM) method to achieve a notable accuracy of 98.27$\%$. It should be noted that this accuracy surpasses the findings of the present work. However, it is important to acknowledge that their experimentation was limited to canines.

\textit{CHB-MIT dataset:} The classification performance is presented in Table \ref{Tab3} and it is compared with the previous works in Table \ref{Tab2}. The data shown in the Table \ref{Tab2} indicates that, as a general trend, solely attention or recurrence-based models performance are poor compared to the ARNN model. This is because the input sequences exhibit both local and global dependencies that are required to detect seizures effectively. Thus, analyzing EEG signals for seizures is a challenge since the attention module struggles to catch local dependencies, while RNN struggles to capture global dependencies in lengthy sequences. In fact, techniques that effectively capture local properties, such as Long Short-Term Memory (LSTM), provide better outcomes ~\cite{Authors31}, but it fails on large segment length. The performance of the proposed model with other existing work is shown in Table \ref{Tab3}. The table indicates that the CNN model loses temporal information, which results in relatively poor performance ~\cite{Authors15}. Conversely, the 1D-MobileNet model exhibits superior performance compared to the CNN model ~\cite{Authors17}. Nevertheless, the proposed ARNN model, which utilizes the self and cross-attention module recurrently, has demonstrated superior performance compared to the baseline and state-of-the-art methods.

\subsection{Comparison with baselines}
The performance of ARNN model was compared with four different baselines. The first baseline is \textit{LSTM} that is widely used for time series data ~\cite{Authors1}.

\textit{LSTM:} This deep learning network excel in capturing temporal dependencies and intricate patterns across EEG channels over time, making them well-suited for tasks like EEG signal classification. However, LSTMs may suffer from vanishing or exploding gradient problems, limiting their ability to model long-range dependencies effectively ~\cite{Authors24, Authors11, Authors27}. \cref{fig7} displays the performance with varying sequence lengths, illustrating the training time and accuracy as a function of sequence length.

\textit{Vision Transformers (ViT):} While not originally designed for EEG data ~\cite{Authors3, Authors28, Authors29}, it can be adapted to process multi-channel EEG segments by treating them as images. The self-attention mechanisms of ViT enable the extraction of global context across channels. However, it looses sequential information of positions, and struggle with capturing fine-grained temporal dynamics ~\cite{ Authors24, Authors11, Authors32}. The loss of fine-grained features leads to poor performance at larger segment lengths for seizure detection, as shown in \cref{fig7}.

\textit{Compact Convolution Transformer (CCT):} It efficiently extracts spatial and temporal features from EEG data using a convolution tokenizer instead of patches is used in ViT ~\cite{Authors32}. CCT helps with tasks like artifact removal but may miss long-range dependencies that are important for EEG analysis. Fig.\ref{fig7a} shows the relative performance of ViT and CCT-2/4 with learnable positional embedding, showing the capability of CCT to extract better features than the ViT.

\textit{R-Transformer:} The Recurrent neural network enhanced transformer uses the local RNN for finding the fine-grained features and transformer for global dependencies but suffers from increased complexity. \cref{fig7b} and \ref{fig7c} shows the training and inference time required compared to baseline methods, and \cref{fig7a} shows drop in performance as sequence length increased. In this work, the defualt values of R-Transformer given at the source location \href{https://github.com/DSE-MSU/R-transformer}{[R-Transformer]} are used.

\subsection{Discussion}
The proposed ARNN model has been compared against four baseline approaches using various segment configurations. The ARNN model consistently provided the superior results. Table \ref{Tab1} provides a detailed breakdown of classification performance for each canine and human participant on the Upenn and Mayo Clinic datasets. The analysis of the table yields several key observations. Firstly, it is apparent that the self-attention-based model, despite demonstrating data parallelism, looses critical local features. \cref{fig7a} illustrates the effects of missing local features in ViT and CCT. This outcome aligns with similar observations found in the case of the CHB-MIT datasets in \cref{Tab2}. The \cref{fig4} provides a comparative analysis of accuracy against epochs of baseline methods with the ARNN model. In evaluating the performance of various models against ARNN, statistical analysis reveals significant differences. An ANOVA test shows a highly significant p-value of $5.36 \times 10^{-12}$, which confirms that the models exhibit different levels of performance. Subsequent pairwise t-tests indicate that CCT (p-value $2.80 \times 10^{-6}$), VT (p-value $0.00186$), and LSTM (p-value $4.19 \times 10^{-7}$) all have statistically significant differences in accuracy compared to ARNN, with ARNN outperforming these models. Interestingly, the RT model, with a p-value of $0.525$, shows no statistically significant difference from ARNN, suggesting that its performance is comparable. The RT model achieves comparable accuracy to ARNN, but it requires significantly longer training and inference time. These findings highlight ARNN's robustness and indicate areas where other models may need further refinement to match its accuracy. The loss of information by these models is shown in Fig. \ref{fig7a}. Notably, the figure demonstrates the effects of sequence length on LSTM and loss of local features in models solely based on attention ~\cite{Authors23}. Sequential models, such as LSTM, typically demanded more time for training and testing compared to attention-based models like ViT and CCT. 

The convolution tokenizer of CCT leveraging convolutional neural networks (CNNs), efficiently dissects the EEG segment into manageable units. By sliding convolutional filters across the EEG signal, it identifies local patterns and extracts meaningful features. However, this approach has drawbacks. Firstly, CNNs lack interpretability, making it challenging to discern token generation rationale. Secondly, fixed filter sizes limit adaptability to diverse bursts of EEG. Thus, while convolution tokenizers offer robust solutions for NLP tasks, their limitations necessitate careful consideration to capture short bursts present in a EEG segment. To overcome these issues, a recurrent neural network enhanced transformer is proposed to sequentially process the sequences of segments by local RNN ~\cite{Authors27}. This R-Transformer is complex and requires high computational time due to sequential process of data using LSTM. To overcome these problems, for time series data, the ARNN model is proposed, where an attention layer is applied recurrently along the sequence length. The ARNN model demonstrates superior performance with shorter sequence lengths, as it effectively captures fine-grained details and dependencies, leading to improved accuracy, as shown in \cref{fig7a}. However, this benefit comes with the trade-off of longer training times, as indicated in \cref{fig7b}. Shorter sequences generate more batches and require more iterations to cover the data set, increasing the overall training time. Conversely, longer sequence lengths reduce training time by requiring fewer batches, but may result in slightly lower accuracy. The ARNN model, while efficient with longer sequences, may not capture all fine-grained details as effectively as it does with shorter sequences, leading to a performance trade-off.

In the experiments, we selected a batch size of 50, which provided a balance between computational efficiency and the accuracy of the gradient updates during training. A smaller batch size would have resulted in noisier gradient estimates, potentially slowing down convergence, while a larger batch size would have required more memory and computation time. To ensure that the choice of batch size did not negatively impact training performance, we compared the training time and convergence behavior with different batch sizes, as shown in Figure \cref{fig6a}. The results demonstrate that a batch size of 50 strikes a good balance, providing efficient training without sacrificing accuracy. Moreover, a dynamic learning rate schedule, starting at 0.001 and decreasing by a factor of 0.1 every 10 epochs, to further optimize training stability and model performance. This approach ensured that the learning process remained effective throughout the training. We have also analyzed the impact of preprocessing on the performance of the ARNN model by comparing the results obtained from filtered and unfiltered CHB-MIT data. Specifically, a second-order Butterworth high-pass filter is used to remove frequencies below 0.5 Hz, followed by a notch filter to eliminate power line interference at 60 Hz and 50 Hz. Interestingly, the ARNN model demonstrated approximately equal performance on both the filtered and unfiltered data, with a statistical significance value of $p=0.92$. This result suggests that the ARNN model is robust to the types of interference typically present in long-term EEG recordings, and the filtering process did not significantly impact the model's accuracy. Thus, the model's inherent design effectively handles noise and interference, minimizing the need for extensive preprocessing. Our qualitative examination, suggests that the model leverages the recurrent state to encapsulate information on commonly encountered features at local locations in EEG signals. Nevertheless, it appears to lack intricate reasoning capabilities at larger time steps but demonstrated superior performance at moderate time steps on fixed segment length configuration, as shown in \cref{fig6}. The LSTM-style gate decides retention or dismissal based on current states and inputs, certain forms of long-range approximate of fine-grained features from the attention mechanisms.

\begin{table}
    \centering
    \rowcolors{3}{gray!15}{gray!5}
    \resizebox{\columnwidth}{!}{%
    \begin{tabular}{c |c c c c c} \cline{2-6}
    \rowcolor{battleshipgrey!80}
         & LSTM& ViT& CCT& RT& ARNN\\ \hline
       Accuracy& 0.661&  0.773& 0.943 & 0.932&0.963\\ \cline{2-6}
         Precision & 0.673 & 0.772 & 0.943 &0.931  & 0.963 \\\cline{2-6}
        Recall & 0.662 & 0.773 & 0.944 &0.932 & 0.962\\\hline   
    \end{tabular}%
    }

    \caption{The accuracy and micro f1-score performance parameter of ARNN on CHB-MIT dataset with the implemented methods at 1024 (4.0 secs) sequence length.}
    \label{Tab2}
\end{table}

\begin{table}[t]
  \centering
  \rowcolors{3}{gray!15}{gray!5}
  \resizebox{\columnwidth}{!}{%
  \begin{tabular}{c | c c c}
    \hline
    \rowcolor{battleshipgrey!80}
        \multicolumn{4}{c}{Upenn and Mayo's clinic}  \\ \hline
        \rowcolor{battleshipgrey!40}
        Study & Year & Method & Accuracy    \\\hline
        Qi \etal ~\cite{Authors42} & 2024 & SSL & 77.24    \\\cline{2-4}
        İlkay \etal ~\cite{Authors18} & 2022 & VAE & 79.00    \\\cline{2-4}
        Nhan \etal ~\cite{Authors19} & 2017 & ACS+ Random forest & 95.81   \\ \cline{2-4}
        Fosil \etal ~\cite{Authors20} & 2020 & SVM & 98.27\\\cline{2-4}
        Nhan \etal ~\cite{Authors30} & 2018 & Integer CNN & 97.10    \\\hline 
        & - & LSTM & 89.37  \\\cline{2-4}
         & - & ViT & 92.09  \\\cline{2-4}
         & - & CCT & 97.95  \\\cline{2-4}
        \multirow{-4}{*}{Baselines}  & - & R-Transformer & 97.82\\ \hline
        Proposed & - & ARNN & 98.20    \\\hline
        
        \rowcolor{battleshipgrey!80}
        \multicolumn{4}{c}{CHB-MIT}  \\ \hline
        Abdulwahhab \etal ~\cite{Authors43} & 2024 & CWT, LSTM & 99.57   \\\cline{2-4}
        Shilpa \etal ~\cite{Authors40} & 2023 & 1D-MobileNet & 98.50   \\\cline{2-4}
        Deepa \etal ~\cite{Authors13} & 2022 & Bi-LSTM & 99.55    \\\cline{2-4}
        
        Fatima \etal ~\cite{Authors17} & 2022 & CNN + ML classifier & 97.10    \\\hline
         & - & LSTM & 98.19  \\\cline{2-4}
         & - & ViT & 97.82  \\\cline{2-4}
         & - & CCT & 98.93  \\\cline{2-4}
        \multirow{-4}{*}{Baselines}  & - & R-Transformer & 99.80\\ \hline
        Proposed & - & ARNN & 99.96    \\\hline
  
  \end{tabular}%
    }
  \caption{Performance comparison of ARNN and baseline methods at sequence length 128 (0.5 secs) for chb-mit and 400 (1.0 secs) for upenn dataset with the state-of-the-art methods on respective datasets.}
  \label{Tab3}
\end{table}

\section{Conclusion}

This study introduces a novel Attentive Recurrent Neural Network (ARNN) architecture designed for efficient and accurate analysis of multichannel EEG data in seizure detection tasks. The ARNN model leverages self-attention and cross-attention layers to extract local interchannel features, which are then processed through an LSTM-style recurrent gate to form global long-term dependencies. This approach allows the model to effectively handle larger EEG segments, a scenario where baseline models often struggle.\par
The effectiveness of the ARNN architecture was validated using heterogeneous datasets, including the CHB-MIT and UPenn \& Mayo Clinic datasets. The results demonstrate that the ARNN outperforms state-of-the-art methods in both accuracy and processing speed. A significant contribution of this study is the careful trade-off between accuracy and computational efficiency. By optimizing the balance between local windows from a larger segment, the ARNN model reduces training time while preserving high detection performance. This trade-off is crucial for practical applications, where rapid and accurate EEG analysis is needed, especially in clinical environments. The empirical findings highlight the potential of the ARNN model to enhance EEG-based clinical interventions, offering a more computationally efficient and precise tool for seizure detection.





\end{document}